\documentclass[sigconf]{acmart}

\AtBeginDocument{%
  \providecommand\BibTeX{{%
    \normalfont B\kern-0.5em{\scshape i\kern-0.25em b}\kern-0.8em\TeX}}}

\setcopyright{acmcopyright}
\copyrightyear{2018}
\acmYear{2018}
\acmDOI{XXXXXXX.XXXXXXX}

\acmConference[Conference acronym 'XX]{Make sure to enter the correct
  conference title from your rights confirmation email}{June 03--05,
  2018}{Woodstock, NY}
\author{Zhizhen Zhang}
\email{zhangzz21@mails.tsinghua.edu.cn}
\affiliation{%
  \institution{Tsinghua University}
  \country{China}
}

\author{Xiaohui Xie}
\email{xiexiaohui@mail.tsinghua.edu.cn}
\affiliation{%
  \institution{Tsinghua University}
  \country{China}
}

\author{Mengyu Yang}
\email{mengyuyang@bupt.edu.cn}
\affiliation{%
  \institution{Beijing University of Posts and Telecommunications}
  \country{China}
}

\author{Ye Tian}
\email{yetian@bupt.edu.cn}
\affiliation{%
  \institution{Beijing University of Posts and Telecommunications}
  \country{China}
}

\author{Yong Jiang}
\email{jiangy@sz.tsinghua.edu.cn}
\affiliation{%
  \institution{Tsinghua University}
  \country{China}
}

\author{Yong Cui}
\authornotemark[1]
\email{cuiyong@tsinghua.edu.cn}
\affiliation{%
  \institution{Tsinghua University}
  \country{China}
}
\renewcommand{\shortauthors}{Zhang, et al.}

\usepackage{caption}
\usepackage{subcaption}
\usepackage{bbding}
\usepackage{multirow}
\usepackage{fancyhdr}
\begin{document}

\title{Improving Social Media Popularity Prediction with Multiple Post Dependencies}

\settopmatter{printacmref=false} 
\renewcommand\footnotetextcopyrightpermission[1]{}
\begin{abstract}
  Social Media Popularity Prediction has drawn a lot of attention because of its profound impact on many different applications, such as recommendation systems and multimedia advertising. Recent work attempts to leverage the content of posts to improve predictive performance. However, few of them consider the multiple dependencies of posts, resulting in their insufficiency to take full advantage of the rich content. To tackle this problem, we propose a novel prediction framework named Dependency-aware Sequence Network (DSN) that exploits both intra- and inter-post dependencies to comprehensively extract content information from posts. For intra-post dependency, DSN adopts a multimodal feature extractor with an efficient fine-tuning strategy to obtain task-specific representations from images and textual information of posts. For inter-post dependency, DSN uses a hierarchical information propagation method to learn category representations that could better describe the difference between posts. DSN also exploits recurrent networks with a series of gating layers for more flexible local temporal processing abilities and multi-head attention for long-term dependencies. The experimental results on the Social Media Popularity Dataset demonstrate the superiority of our method compared to existing state-of-the-art models.
\end{abstract}



\maketitle
\section{Introduction}
Social media is an essential part of people's lives. Understanding the content of social media and forecasting its popularity has drawn a lot of attention from researchers both in academia and industry\cite{pinto2013using,khosla2014makes,he2014predicting}. Precise popularity prediction can greatly benefit various applications, such as online advertising, content recommendation, and trend analysis. In this paper, we focus on the Social Media Popularity Prediction (SMPP) task, which aims to estimate the target post's future popularity via plenty of social media data. Typically, the data includes the post content (e.g., images and textual description) and the information of the user who posted it. Among them, user information is usually numerical and easier to process (e.g. number of followers, number of likes), while content information is more complicated and also an important factor for users to interact and share on social media platforms. Engaging content can attract more views, likes, and shares, leading to increased popularity. Conversely, unattractive content is less likely to be engaged with and shared, resulting in less popularity. Therefore, accurately predicting the popularity of a post requires a comprehensive understanding of its content. Conventional SMPP works often manually extract image and textual features, concatenating them directly and then applying machine learning algorithms to make 
prediction\cite{wang2017combining,he2019feature,kang2019catboost}. This approach does not take full advantage of the post content, resulting in poor predictive performance.\par 

Recently, to make better use of the multimodal content, some works consider the correlation between different modalities within a post, which we called intra-post dependency. Among them, Xu et al. introduce an attention mechanism to assign large weights to more important modalities\cite{xu2020multimodal}. Some others apply multimodal learning methods to align different modalities\cite{tan2022efficient, chen2022and}. However, the inter-post dependency, on the other hand, capturing the association between different posts, is not well modeled by their work. For example, posts from the same user or on the same topic may share similar content or attract similar audiences over a sustained period, aggregating information from relevant posts might also enhance the post representation.\par 

In that regard, many researchers adopt sequence modeling to enhance the representation of the target post by incorporating temporally correlated posts. Among them, some extract temporal features of the target user's posting sequence with sliding window moving average or temporal transformer.\cite{wang2020feature, tan2022efficient}, but the methods based on user post sequence ignore the correlation across different users. Wu \textit{et al.} use the recurrent network and temporal attention mechanisms to model temporal coherence of posts from different users across multiple time-scales\cite{wu2017sequential}. However, the temporal attention mechanism only considers the release time, other context information is poorly modeled.\par

\begin{figure}[t]
    \centering
    \includegraphics[width=0.3\textwidth]{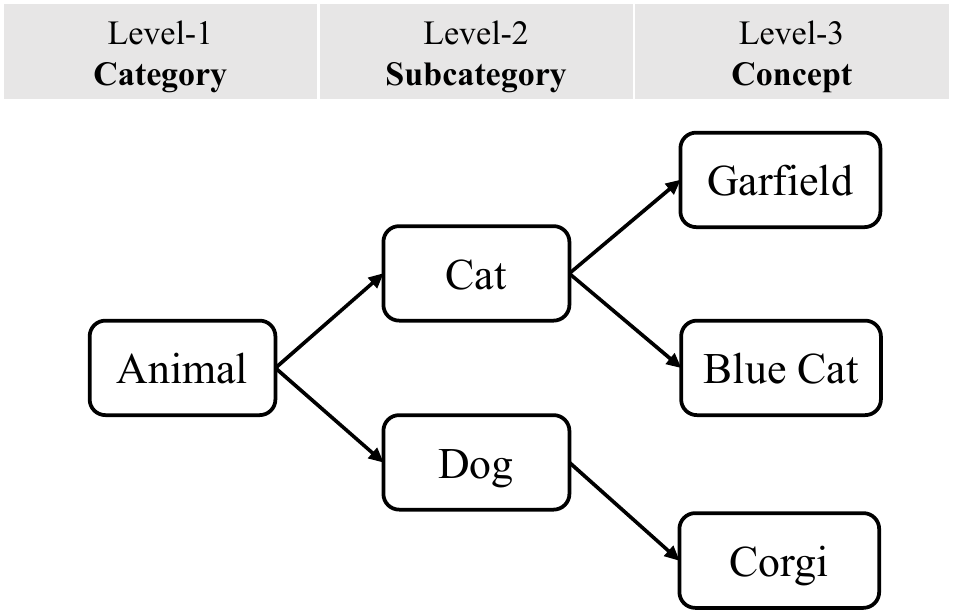}
    \caption{An example of three-level hierarchical category information. From left to right, three levels, i.e., Category, Subcategory, and Concept are presented. }
    \label{fig:category}
\end{figure}

Besides the temporal dependency, the categories of posts are also important for modeling the correlation between different posts, which are not well considered by previous works. Posts on social platforms are tagged when they are published to describe the category of the content. These tags are usually hierarchical, for example, photos of the sky could be tagged with landscape, nature, etc to get a higher probability of being viewed. This hierarchy enables category information not only to describe the content of a single post but also to model the correlation between different posts in a fine-grained manner. Figure \ref{fig:category} shows an example of three-level hierarchical category information. The correlation between Garfield and Blue Cat is closer than between it and Corgi, because the former two belong to cats, while the latter belongs to dogs, although they both belong to animals. If we only use the category embedding of a single level, the difference between them could not be accurately modeled.\par

In this paper, we propose a novel deep popularity prediction framework called Dependency-aware Sequence Network (DSN) that leverages both intra- and inter-post dependencies to comprehensively extract content information from posts, which could help us gain a more comprehensive understanding of the factors that contribute to post popularity and improve the accuracy of popularity prediction. We design the architecture to be consistent with multimodal inputs and temporal relationships common to social media popularity prediction - specifically incorporating (1) multimodal feature extractors which explore the correlation between images and textual information of posts. (2) a hierarchical modeling approach that exploits the hierarchical nature of category information to improve post-similarity modeling. (3) a sequence-to-sequence layer with multi-head attention to aggregate post inputs. A series of gating layers are conducted to give the model more flexibility to model the dependencies between posts.\par
Overall, our contributions can be summarized as follows:
{\begin{itemize}
    \item We fully leverage the importance of the post content for social media popularity prediction by formally defining and modeling intra- and inter-post dependencies.
    \item We present a novel prediction model called DSN that achieves more precise popularity prediction by jointly modeling the correlation between images and text, the hierarchy of category information, and temporal relevance between posts.
    \item We conduct extensive experiments to investigate the effectiveness of DSN. The experimental results verify the efficacy and superiority of DSN over state-of-the-art models on Social Media Popularity Dataset. For the convenience of the reproduction of the results, we will make our code publicly available upon publication.
\end{itemize}

\begin{figure*}[!htbp]
    \centering
    \includegraphics[width=1.0\textwidth]{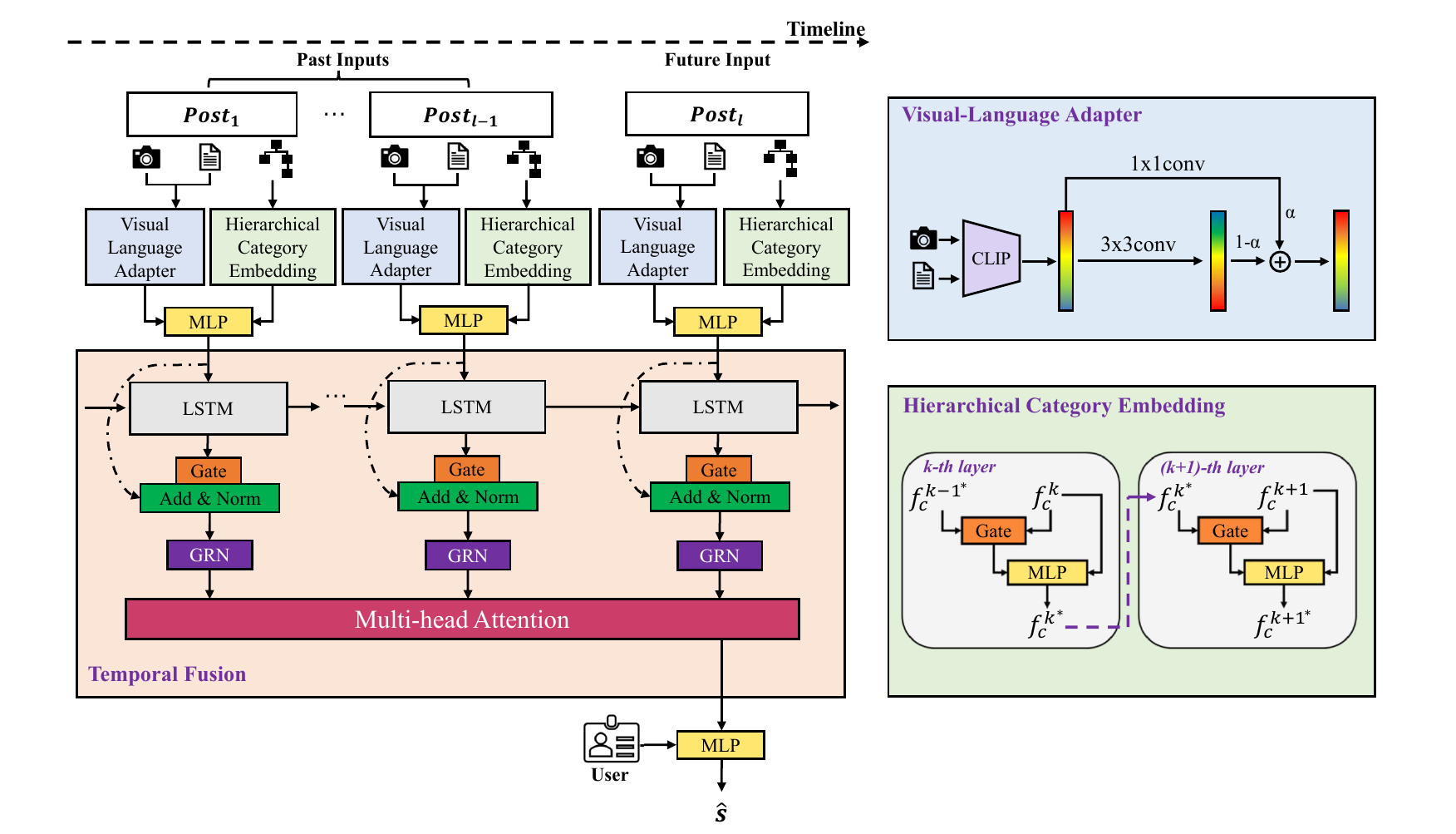}
    \caption{Overview of the proposed model DSN. DSN inputs various content information (i.e. image, text, category) of the post sequence. Visual-Language Adapter (blue region) is used for extracting task-specific multimodal features from images and text. Hierarchical Category Embedding (green region) incorporates the category information across different levels for better describing the difference between posts. Temporal Fusion (orange region) is based on LSTM for local processing and multi-head attention for integrating information at any time step in the post sequence. At last, the enhanced target post feature is fused with user information for the final prediction.}
    \label{fig:model}
\end{figure*} 

\section{Related Work}
For the SMPP task, conventional works manually extract image and textual features of posts, fusing them with other metadata (\textit{e.g.} user information) and make predictions with regression models. \cite{chen2019social, ding2019social}. However, these efforts fail to take full advantage of the posts' rich content information. Recently, many works start to consider the relationship between different modalities, which we denote as intra-post dependency, to enhance the post representation. Among them, Xu \textit{et al.} propose a multimodal deep learning framework that introduces an attention mechanism to assign large weights to specified modalities\cite{xu2020multimodal}.
Chen \textit{et al.} build two-stream ViLT models for title-visual and tag-visual representations, and design title-tag contrastive learning for two streams to learn the differences between titles and tags\cite{chen2022and}.
Tan \textit{et al.} first perform visual and textual feature extraction respectively and then employ a multimodal transformer ALBEF to align visual and text features in semantic space\cite{tan2022efficient}. These models only consider the interaction between modalities within a single post, neglecting the correlation between different posts.
\par
To solve this problem, some works target to utilize the temporal correlation between different posts, which is among inter-post dependency, to enhance the feature representation of the post to be predicted. Among them, Wu \textit{et al.} analyze temporal characteristics of social media popularity, consider the posts as temporal sequence and make a prediction with temporal coherence across multiple time-scales\cite{wu2017sequential}. 
Wang \textit{et al.} use sliding window average to mine potential short-term dependency for each user’s post sequence, then predict by a combined Catboost model to handle the problem of data missing\cite{wang2020feature}.
Tan \textit{et al.} propose a transformer and sliding window average-based timing feature extraction method to reduce the inconsistent distribution between timing features extracted from the training set and test set\cite{tan2022efficient}. However, these works only concentrate on temporal modeling while ignoring other contextual information. There is not only a temporal relationship between posts, the category difference in post content can also describe the correlation between posts.\par

In summary, few of existing works jointly model multiple dependencies of posts, leading them to fail to exploit the potential of post content in social media popularity prediction. They either do not consider the relevance of different modalities of the single post or ignore the correlation between posts from different users or other context information (\textit{e.g.} hierarchical category information). To overcome these problems, our work provides a comprehensive overview of multiple dependencies of posts from both inter- and intra-post perspectives to fully model the content information. 

\section{PROPOSED METHOD}
\subsection{Problem Definition}
Formally, given a new post $p$ published by user $u$, the problem of predicting its popularity is to estimate how much attention it would receive after its release (\textit{e.g.} views, clicks or likes \textit{etc.}). In Social Media Popularity Dataset\cite{wu2019smp} which we use for the experiment,  “viewing count” is used to describe the popularity after log-normalization as below:
\begin{equation}
s=\log _2 \frac{r}{d}+1
\end{equation}
where $r$ is the viewing count, $d$ is the number of days since the photo was posted and $s$ is the normalized popularity.\par
In this paper, for any given post $p_i$ released at timestamp $t_i$, we adopt a chronological sliding window with length $l$ to get the post sequence, which can be defined as $P_i = (p_{i-l+1}, p_{i-l+2},..., p_{i-1}, p_i)$, where $t_{i-l+1} < t_{i-l+2} < ... < t_{i}$. We aim to build a model $F$ which can generate a popularity score $\hat{s_i}$ from $P_i$ for the post $p_i$:
\begin{equation}
\hat{s_i} = F(P_i)
\end{equation}

\subsection{Overview of DSN}
We design DSN by using canonical components to efficiently build feature representations for rich contents of posts (i.e. image, text, category) to obtain high predictive performance. The major constituents
of DSN are:\par
1. \textbf{Visual-Language Adapter} to extract visual and textual information, employing a multimodal pre-trained model with an efficient fine-tuning strategy to get task-specific representations.\par
2. \textbf{Hierarchical Category Embedding} to use a gating mechanism to permit the valuable category information to be passed from coarse to fine granularity. The obtained representation incorporates the information at different levels and can better describe the correlation between posts with different categories.\par
3. \textbf{Temporal Fusion} to learn both short- and long-term temporal relationships from past posts. A sequence-to-sequence layer is employed for local processing, whereas long-term dependencies are captured using a multi-head attention block. Gating mechanisms are also used to skip over any unused components of the architecture, providing adaptive depth and network complexity.\par
The overall framework of DSN is shown in Figure \ref{fig:model}, with individual components described in detail in the subsequent sections. 
\subsection{Visual-Language Adapter}
Visual-Language Adapter is used to generate a unified representation of visual and textual descriptions of the posts. There has been a lot of vision-language pre-trained models exploring the interaction between these two modalities, one of them is CLIP (Contrastive Language-Image Pretraining)\cite{radford2021learning} which achieves astonishing results on a wide range of vision tasks without any fine-tuning.  To close the gap between CLIP and downstream popularity prediction task, inspired by CLIP-Adapter\cite{gao2021clip}, we design an efficient feature adapter that only appends a small number of additional learnable layers with residual connections to CLIP’s language and image branches while keeping the original CLIP backbone frozen during fine-tuning.\par
Specifically, given the input image $I$ and textual description $T$ of the post sequence $P$, the original visual and textual embedding $\mathbf{f}_v^{origin}, \mathbf{f}_t^{origin}\in \mathbb{R}^{l\times d_{origin}}$ are computed with CLIP backbone, where $l$ is the sequence length and $d_{origin}$ is the dimension of the output of the CLIP encoder. After that, two learnable feature adapters $A_v(\cdot)$ and $A_t(\cdot)$ are adopted to transform $\mathbf{f}_v^{origin}$ and $\mathbf{f}_t^{origin}$, respectively. In each adapter, we use convolutions with 3x3 filters and ReLU activation function\cite{glorot2011deep} to get adapted features, and convolutions with 1x1 filters to reserve the original knowledge encoded by CLIP. For each convolutional layer, we perform downscaling from $d_{origin}$ to $d_{hidden}$ so that these features could be consistent with other low-dimensional features, eg., the user information of the post). Two trade-off parameters $\alpha$ and $\beta$ are employed as “residual ratio” to help adjust the degree of maintaining the original knowledge for better performance. In summary, for given input $\mathbf{x}$, the feature adapters can be written as:
\begin{equation}
    A_v(\mathbf{x}) = (1-\alpha) {{\rm ReLU(Conv}_{3{\rm x}3}(\mathbf{x}))}  + \alpha {{\rm Conv}_{1{\rm x}1}(\mathbf{x})}
\end{equation}
For $A_t(x)$, $\alpha$ is replaced with $\beta$. After employing adapters to the original features, we can get new visual and textual feature $\mathbf{f}_v, \mathbf{f}_t \in \mathbb{R}^{l\times d_{hidden}}$:
\begin{align}
    \mathbf{f}_v &= A_v(\mathbf{f}_v^{origin}) \\
    \mathbf{f}_t &= A_t(\mathbf{f}_t^{origin})
\end{align}
\subsection{Hierarchical Category Embedding}
Hierarchical Category Embedding is to learn category representation which can better describe the correlation between different posts by stacking several HCE layers. Each layer uses a gating mechanism that allows valuable information from the previous layer to pass through and combine with information from this layer to obtain a comprehensive category representation across layers.\par
Specifically, for the hierarchy category information as shown in figure \ref{fig:category}, we denote the original category embedding of level-$k$ learned by the embedding layer as $\mathbf{f}_c^{k} \in \mathbb{R}^{l \times h_{hidden}}$. The inputs of the $k$-th HCE layer are the independent category embedding $\mathbf{f}_c^{k}$ of level-$k$ and the hierarchical embedding $\mathbf{f}_c^{{k-1}^*}$ computed by the last layer. The output $\mathbf{f}_c^{{k}^*}$ of the $k$-th layer would be treated as the input to the next layer for iterative computing:
\begin{equation}\label{hce}
    \mathbf{f}_c^{k^{*}} = {\rm HCE}_k\left(\mathbf{f}_c^{{k-1}^*}, \mathbf{f}_c^{k}\right)
\end{equation}

In each HCE layer, we use a gating mechanism based on Gated Linear Unit  (GLU)\cite{dauphin2017language} to compress the information from the previous layer. The more valuable the information of the previous layer is, the more it would be retained. Given input $\mathbf{x}, \mathbf{y} \in \mathbb{R}^{l \times d_{hidden}}$, our gating mechanism can be represented as:
\begin{equation}
    {\rm Gate}_k (\mathbf{x}, \mathbf{y}) = (\mathbf{x}\mathbf{W}_{1,k} + \mathbf{b}_{1,k}) \odot \sigma(\mathbf{x}\mathbf{W}_{2,k} + \mathbf{y}\mathbf{W}_{3,k} + \mathbf{b}_{2,k}) \label{gate}
\end{equation}
where $\mathbf{W}_{(.)}\in \mathbb{R}^{d_{hidden}\times d_{hidden}}, \mathbf{b}_{(.)}\in \mathbb{R}^{d_{hidden}}$ are the weights and bias, $k$ means the parameters of the gating mechanism are not shared across different HCE layers. $\odot$ is the element-wise Hadamard product, and $\sigma(\cdot)$ is the sigmoid activation function.\par
With the gating mechanism, we can give the calculation process of the HCE layer in Eq. \ref{hce}:
\begin{align}
    \mathbf{f}_c^{{k-1}^{\prime}} &= {\rm Gate}_k (\mathbf{f}_c^{{k-1}^*}, \mathbf{f}_c^{k})\label{gate_hce} \\
    \mathbf{f}_c^{{k}^*} &= \left[\mathbf{f}_c^{k} \| \mathbf{f}_c^{{k-1}^\prime}\right]\mathbf{W}_k\label{layer_fuse}
\end{align}
where $\mathbf{W}_{k}\in \mathbb{R}^{2d_{hidden}\times d_{hidden}}$ is the weight of the $k$-th HCE layer and $\|$ is the concatenation operation. Eq. \ref{gate_hce} generates a gating of cross-layer information through a sigmoid function to control the inflow of relevant information from the previous layers. A linear transformation is used in Eq. \ref{layer_fuse} to combine the useful information from previous layers with the independent information of the current layer.\par
We stack three HCE layers to model the 3-level hierarchy category information given by the dataset. The output $\mathbf{f}_c^{{3}^*}$ of the last HCE layer incorporates category information across different levels and can express the differences between posts in a more fine-grained manner. We denote $\mathbf{f}_c^{{3}^*}$ as $\mathbf{f}_c$ for a unified description. Then $\mathbf{f}_c$ is concatenated with visual-language features computed by Visual-Language Adapter as the post representations $\mathbf{f}\in \mathbb{R}^{l\times 3d_{hidden}}$ of the post sequence:
\begin{equation}
    \mathbf{f} = \left[\mathbf{f}_v || \mathbf{f}_t || \mathbf{f}_c\right]
\end{equation}

\subsection{Temporal Fusion}
Temporal Fusion is to learn temporal dependency between posts. Note that $P_l$ might contains posts from several different users, which presents a challenge for the temporal modeling of post sequences. We construct the module based on LSTM which is commonly used for local processing. Inspired by Temporal Fusion Transformer\cite{lim2021temporal}, we use the gating mechanism and residual connection to control the fusion of the temporal information learned by LSTM and the original features.
Specifically, given the input $\mathbf{f}\in \mathbb{R}^{l\times 3d_{hidden}}$, the output $\mathbf{\Phi} \in \mathbb{R}^{l\times d_{hidden}} $ of local temporal processing can be calculated as follow:
\begin{align}
    \mathbf{O} &= {\rm LSTM}(\mathbf{f}) \\
    \mathbf{\Theta} &= {\rm LayerNorm(Gate}_o(\mathbf{O}) + \mathbf{f}\mathbf{W}) \label{reslstm} \\
    \mathbf{\Phi} &= {\rm GRN}_{\theta}(\mathbf{\Theta}) \label{grn}
\end{align}
where LayerNorm is standard layer normalization of \cite{ba2016layer}, and $\mathbf{W} \in \mathbb{R}^{3d_{hidden}\times d_{hidden}}$ is the linear transformation for downscaling. The gating mechanism is a version of Eq.\ref{gate} with only one input. $o$ denotes the parameters of the gating mechanism here are shared across the entire layer. The Gated Residual Network (GRN)\cite{lim2021temporal} in Eq.\ref{grn} could provide adaptive depth to the model. Given an input $\mathbf{x}$, the GRN yields:
\begin{align}
    {\rm GRN}_{\theta}(\mathbf{x}) &= {\rm LayerNorm}(\mathbf{x} + {\rm Gate_{\theta}(\mathbf{\eta}_1)}) \\
    \mathbf{\eta}_1 &= \mathbf{W}_{1,\theta} \mathbf{\eta}_{2} + \mathbf{b}_{1,\theta} \\
    \mathbf{\eta}_2 &= {\rm ELU}(\mathbf{W}_{2,\theta} \mathbf{x} + \mathbf{b}_{2,\theta})
\end{align}
where ELU is the Exponential Linear Unit activation function\cite{clevert2015fast}. The gating mechanism and residual structure in GRN could provide flexibility to apply non-linear processing when needed. If necessary, the layer could be entirely skipped, as the outputs of Gate may all be close to zero to suppress nonlinear contributions.\par
We further adopt multi-head attention\cite{vaswani2017attention} to learn long-term dependency in the post sequence. Here, the attention mechanism computes dot-product attention, which is defined as:
\begin{equation}
    {\rm Attetion}(\mathbf{Q, K, V}) = {\rm Softmax}(\mathbf{Q}\mathbf{K}^{T}/\sqrt{d_{hidden}})\mathbf{V}
\end{equation}
where $\mathbf{Q}$ denotes the 'query', $\mathbf{K}$ the 'key' and $\mathbf{V}$ the 'value'. While the usual sequence-sequence model calculates the attention between any two positions, our model DSN is concerned with the correlation between the target post and other posts in the sequence. Given the input $\mathbf{\Phi}$ yields by Eq.\ref{grn}, the 'query', 'key' and 'value could be obtained as below: 
\begin{equation}
    \mathbf{q} = \mathbf{\Phi}_{l}\mathbf{W}^Q, \mathbf{K} = \mathbf{\Phi}_{1:l-1} \mathbf{W}^K, \mathbf{V} = \mathbf{\Phi}_{1:l-1} \mathbf{W}^V 
\end{equation}
where $\mathbf{W}^Q, \mathbf{W}^K, \mathbf{W}^V \in \mathbb{R}^{d_{hidden}\times d_{hidden}}$ are the weight matrices. $\mathbf{\Phi}_{l}\in\mathbb{R}^{d_{hidden}}$ is the feature of the target post and $\mathbf{\Phi}_{1:l-1}\in\mathbb{R}^{(l-1) \times d_{hidden}}$ are of other posts.  We then use the above dot-product attention to get the hidden representations of the neighbor posts $\mathbf{h}\in\mathbb{R}^{d_{hidden}}$:
\begin{equation}
\mathbf{h} = {\rm Attetion(} \mathbf{q}, \mathbf{K}, \mathbf{V}) \\
\end{equation}
To combine neighbor post representations $\mathbf{h}$ with the target post feature $\mathbf{\Phi}_{l}$, we concatenate them and feed them into a feed-forward neural network to capture non-linear interactions between the features as in \cite{vaswani2017attention}:
\begin{equation}
    \mathbf{\tilde{h}} = {\rm FFN} \left(\mathbf{h} || \mathbf{\Phi}_{l}\right) = \rm{ReLU}(\left[ \mathbf{h} || \mathbf{\Phi}_{\textit{l}}\right] \mathbf{W}_1 + \mathbf{b}_1 )\mathbf{W}_2 + \mathbf{b}_2
\end{equation}
where $\mathbf{W}_1 \in \mathbb{R}^{2d_{hidden} \times d_{hidden}}$, $\mathbf{W}_2 \in \mathbb{R}^{d_{hidden} \times d_{hidden}}, \mathbf{b}_1, \mathbf{b}_2 \in \mathbb{R}^{d_{hidden}}$, and $\mathbf{\tilde{h}} \in \mathbb{R}^{d_{hidden}}$ is the final output representing the target post feature.
\subsection{Prediction Module}
Considering the importance of user information shown in previous works, we choose some metadata from the official Social Media Popularity Dataset. We also follow the HyFea\cite{lai2020hyfea} method to obtain more numerical information from users' websites. All the user information we use is listed in Table \ref{tab:metadata}. For categorical data Uid, we adopt a learnable embedding layer to transform it into numerical features. For PublishTime, we convert the timestamp to month, day and hour and process them by one-hot encoding. The rest of the data are all numerical and we scale them by z-score normalization. All information about the user who posted the target post $p$ is concatenated as user features $\mathbf{u}$. After concatenating user features $\mathbf{u}$ and post features $\mathbf{\tilde{h}}$, a two-layer MLP (Multi-Layer Perception) is adopted to learn the relationship between the user and post and perform the popularity prediction:
\begin{equation}
    \hat{s} = {\rm MLP}(\left[\mathbf{\tilde{h}} || \mathbf{u}\right])
\end{equation}
Finally, we minimize the MSE (Mean Square Error) loss to optimize the predicted popularity value $\hat{s}$.
\begin{table}
    \centering
    \begin{tabular}{lrr}
        \toprule
        Data Entry &  Description\\
        \midrule
        Uid & The user this post belongs to.\\
        Ispublic &  Is the post authenticated with 'read' permissions.\\
        Ispro & Is the user belong to pro member.\\      
        Latitude & The latitude of the posting location.\\ 
        Longitude & The longitude of the posting location.\\
        GeoAccuracy & The accuracy level of the location information.\\
        Postdate & The publish timestamp of the post.\\
        \midrule
        $\rm{Followers^*}$ & The number of people the user follows.\\
        $\rm{Following^*}$ & The number of followers of the user.\\
        $\rm{Views^*}$ & The number of views of the user's posts.\\
        $\rm{Tags^*}$ & The number of tags of the user's posts.\\
        $\rm{Faves^*}$ & The number of faves of the user's posts.\\
        $\rm{InGroups^*}$ & The number of groups the user belongs to.\\
        \bottomrule
    \end{tabular}
    \caption{The user information used for prediction. The data entries with $*$ are crawled from the users' homepages according to Hyfea's method\cite{lai2020hyfea}, and others are from the official SMPD\cite{wu2019smp}.}
    \label{tab:metadata}
\end{table}

\section{Experiment}
\subsection{Dataset} We use the Social Media Prediction Dataset (SMPD)\cite{wu2019smp} collected from Flickr, which is widely used by previous works\cite{ding2019social, xu2020multimodal, kang2019catboost, tan2022efficient, chen2022and}, to evaluate the performance of our method. SMPD contains 486k posts from 69k users. For each post, both visual and textual information are provided along with multiple metadata and category information. The category information is depicted in a 3-level manner. The number of categories in the three levels is 11, 77, and 668, respectively. These posts are sorted by posting time, split for train and test by ratio 2:1. The labeled training set is released to participants, and the labels of the test set for final evaluation have not been released. We use the labeled training set to evaluate our algorithm. We split the data chronologically based on the time order of posts. The ratio of the split is 8:1:1, meaning that 80\% of the data is used for training, 10\% for validation, and 10\% for testing. 

\subsection{Implementation Details} We use CLIP\cite{radford2021learning} as the backbone of Visual-Language Adapter. The dimensions of all hidden layers in the model are set to 256. The number of attention heads is set to 4. The length of the inputted post sequence is set to 16. We optimize the model by Adam optimizer\cite{Kingma2014AdamAM} with the learning rating of 1e-3 and weight decay of 1e-4 for 10 epochs. The batch size is 512. To avoid over-fitting, we set dropout to 0.25. The experiments are implemented with PyTorch and conducted on a single NVIDIA GTX 1080 GPU.

\subsection{Evaluation Metrics} To evaluate the prediction performance, we use a precision metric Mean Absolute Error (MAE) and a correlation metric Spearman Ranking Correlation (SRC) as in \cite{wu2017sequential}. If there are k samples, given ground-truth popularity set $S$ and predicted popularity set $\hat{S}$ varying from 0 to 1, the MAE can be expressed as:
\begin{equation}
M A E=\frac{1}{k} \sum_{i=1}^n\left|\hat{S}_i-S_i\right|
\end{equation}
the SRC is used to measure the ranking correlation between $\hat{S}$ and $S$:
\begin{equation}
S R C=\frac{1}{k-1} \sum_{i=1}^k\left(\frac{S_i-\bar{S}}{\sigma_S}\right)\left(\frac{\hat{S}_i-\overline{\hat{S}}}{\sigma_{\hat{S}}}\right)
\end{equation}
Lower MAE / higher SRC refers to better performance.

\subsection{Baselines}
To showcase the effectiveness of the proposed model, we compare its prediction performance with six state-of-the-art baseline models of SMPP. We summarize dependencies used by different models in Table \ref{tab:comparison} for comparison.\par
\textit{Baseline 1}: \textbf{Deep Context Neural Network (DTCN)\cite{wu2017sequential}}. Wu \textit{et al.} use ResNet\cite{he2016deep} to generate visual representation, jointly embedding them with user feature into in a common space. Based on the embedded sequence over time, they adopt LSTM\cite{hochreiter1997long} and temporal attention to predict popularity with temporal coherence across multiple time scales.\par

\textit{Baseline 2}: \textbf{Multiple Layer Perceptron (MLP)\cite{ding2019social}}. Ding \textit{et al.} use ResNet\cite{he2016deep}, NIMA\cite{talebi2018nima} and IIPA\cite{ding2019intrinsic} to generate deep visual representations, aesthetics scores, and intrinsic popularity scores, respectively. They adopt BERT\cite{devlin2018bert} to get text feature and feed them together with user feature into MLP to make prediction.\par

\textit{Baseline 3}: \textbf{MLP with Attention Mechanism (Att-MLP)\cite{xu2020multimodal}}. Xu \textit{et al.} adopt ResNet\cite{he2016deep} and Word2Vec\cite{mikolov2013efficient} for visual and text respectively. They consider specific modalities are of greater importance on the popularity of the post, using an attention mechanism to control how much attention should be attended to each modality.\par

\textit{Baseline 4}: \textbf{Feature Generalization Framework with Combined Catboost (Catboost)\cite{kang2019catboost}}. Kang \textit{et al.} adopt ResNet\cite{he2016deep} for visual features. They use BERT\cite{devlin2018bert}, FastText\cite{joulin2016bag}, TFIDF, and LDA\cite{blei2003latent} to get text features. They do sliding window moving average over temporal ordered features to model dependency for each user’s posts.\par

\textit{Baseline 5}: \textbf{Efficient Multi-View multimodal Data Processing Framework (Multi-view)\cite{tan2022efficient}}. Tan \textit{et al.} use ALBEF\cite{li2021align} which consists of an image encoder, a text encoder, and a multimodal encoder to extract visual-language representation features. They also use sliding window moving average and transformer based-methods, including Performer\cite{choromanski2020rethinking} and Linformer\cite{wang2020linformer} to extract temporal features for each user's posts.\par

\textit{Baseline 6}: \textbf{Title-and-Tag Contrastive Vision-and-Language Transformer (TTC-VLT)\cite{chen2022and}}. Chen \textit{et al.} use pre-trained ViLT\cite{kim2021vilt} to extract both image and text features. To tackle the problems caused by the difference between titles and tags, they build 2 two-stream ViLT models for title-visual and tag-visual, exploiting contrastive learning to estimate a lower bound of the mutual information between titles and tags.\par

\begin{table}
    \centering
    \begin{tabular}{lcccc}
        \toprule
        \multirow{2}{*}{Methods} & \multicolumn{2}{c}{Dependencies} & \multirow{2}{*}{MAE$\downarrow$} & \multirow{2}{*}{SRC$\uparrow$} \\
         & intra-post & inter-post & & \\
        \midrule
        DTCN\cite{wu2017sequential} & $/$ & $\mathrm{T}$ & 1.532 & 0.624 \\
        MLP\cite{ding2019social} & $/$ & $/$ & 1.483 & 0.631 \\
        Attention MLP\cite{xu2020multimodal} & $\mathrm{V}$-$\mathrm{L}$ & $/$ & 1.453 & 0.635 \\ 
        Catboost\cite{kang2019catboost} & $/$ & $\mathrm{T}$ & 1.442 & 0.663 \\ 
        Multi-view\cite{tan2022efficient} & $\mathrm{V}$-$\mathrm{L}$ & $\mathrm{T}$ & 1.387 & 0.693 \\
        TTC-VLT\cite{chen2022and} & $\mathrm{V}$-$\mathrm{L}$ & $/$ & 1.346 & 0.711 \\
        \midrule
        DSN (Ours) & $\mathrm{V}$-$\mathrm{L}$ & $\mathrm{T}$ \textbf{\&} $\mathrm{C}$ & $\mathbf{1.192^*}$ & $\mathbf{0.763^*}$ \\
        \bottomrule
    \end{tabular}
    \caption{Overall comparison results with state-of-the-art methods on SMPD dataset. Lower MAE / higher SRC refers to better performance. The post dependencies utilized by the methods are listed by intra- and inter-post. V-L denotes visual-language, T denotes temporal and C denotes category. A paired t-test is performed and $*$ indicates a statistical significance $p < 0.001$ compared to the best baseline method. The best results are in bold.}
    \label{tab:comparison}
\end{table}

\subsection{Overall Performance}
We report the best prediction results of our proposed method and the compared models in Table \ref{tab:comparison}. Overall, our model achieves the best prediction performance with the minimal MAE of 1.192 and the highest SRC of 0.763. Compared with the strongest baseline model, \textit{i.e.} TTC-VLT\cite{chen2022and}, our method reduces MAE by 11.0\% and improves SRC by 7.3\%.\par 

From the perspective of the dependencies, Attention MLP\cite{xu2020multimodal} and Catboost\cite{kang2019catboost} utilize intra-post and inter-post dependencies, respectively, making them superior to MLP\cite{ding2019social}. However, the quality of features is also important, e.g., although DTCN\cite{wu2017sequential} exploits the temporal dependency between posts, it extracts insufficient visual features and does not utilize textual features, which leads to unsatisfied performance. Multi-view\cite{tan2022efficient} both consider intra- and inter-post dependencies, making them more effective than the above baselines. TTC-VLT\cite{chen2022and} considers the relevance of image and different types of textual information, i.e. titles and tags, which makes up for the fact that it does not exploit the inter-post dependency.\par
Overall, besides image-text dependency and temporal dependency, our method further considers the hierarchical nature of categories, allowing for more fine-grained modeling of inter-post dependency. Jointly learning both intra- and inter-post dependency makes our model achieve the best prediction result.

\begin{table}
    \centering
    \begin{tabular}{lrrrrrr}
        \toprule
        $l$ & 1 & 4 & 8 & 16 & 32 & 64 \\
        \midrule
        MAE$\downarrow$ & 1.245 & 1.221 & \textbf{1.192} & 1.198 & 1.217 & 1.232\\
        SRC$\uparrow$ & 0.743 & 0.759 & \textbf{0.763} & \textbf{0.763} & 0.758 & 0.747\\
        \bottomrule
    \end{tabular}
    \caption{Influence of sequence length. $l=8$ is the length used by DSN. The best results are in bold.}
    \label{tab:length}
\end{table}

\begin{table}
    \centering
    \begin{tabular}{ccccrr}
        \toprule
        User & Image & Text & Category & MAE$\downarrow$ & SRC$\uparrow$ \\
        \midrule
        \Checkmark & \XSolidBrush & \XSolidBrush & \XSolidBrush & 1.350 & 0.681 \\   
        \Checkmark & \Checkmark & \XSolidBrush & \XSolidBrush & 1.304 & 0.717 \\
        \Checkmark & \XSolidBrush & \Checkmark & \XSolidBrush & 1.303 & 0.707 \\
        \Checkmark & \XSolidBrush & \XSolidBrush  & \Checkmark & 1.269 & 0.727 \\
        \Checkmark & \Checkmark & \Checkmark & \XSolidBrush & 1.289 & 0.723 \\
        \Checkmark & \Checkmark & \XSolidBrush & \Checkmark & 1.215 & 0.752 \\
        \Checkmark & \XSolidBrush & \Checkmark & \Checkmark & 1.231 & 0.741 \\
        \Checkmark & \Checkmark & \Checkmark & \Checkmark & \textbf{1.192} & \textbf{0.763} \\
        \bottomrule
    \end{tabular}
    \caption{Ablations on different combinations of features. The last line is the features used by DSN. The best results are in bold.}
    \label{tab:feature}
\end{table}

\subsection{Ablation Study}
To further illustrate the advantages of the proposed model, we also conduct ablation studies to evaluate the contribution of each module.

\subsubsection{Ablating the sequence length $l$} We experiment on sequence length $l$ to find how much past inputs are best for predicting the target post. From Table \ref{tab:length}, we can see the result is worst when $l=1$, because the model makes predictions only by the information of the target post, ignoring the inter-post dependency. As the sequence length gets longer, the best results are obtained at $l=8$. The results at $l = 16$ are close to optimal. Then as $l$ continues to increase, the results start to get worse. The post sequence of appropriate length facilitates the model to learn inter-post dependency, which improves the prediction performance compared to using only the feature of the target post itself. However, the too-long sequence may incorporate irrelevant information, making it difficult for the model to capture the correct correlation, thus compromising the prediction results.

\subsubsection{Contributions of Different Features} \label{section:feature} In order to better understand the contribution of multimodal content (image, text and category features) to the prediction performance, we take user information as the basic feature and add different combinations of the other three features to evaluate the model performance. Table \ref{tab:feature} shows the results of the ablation study, from which we find that each feature improves the result to a certain extent, and more features produce better prediction performance. In terms of the effect of different features, the category information improves the results more than images and text. The reason might be that compared with the abstract semantics contained in text and images, category information can more accurately represent whether a post belongs to the popular type. The text data improves the performance least, and the reason might be that the text data in SMPD contains a large amount of semantically ambiguous expressions (\textit{e.g.} meaningless sequence of numbers or abbreviation of words), which could make it difficult for the model to learn useful textual information.

\begin{figure}[t]
    \centering
    \includegraphics[width=0.5\textwidth]{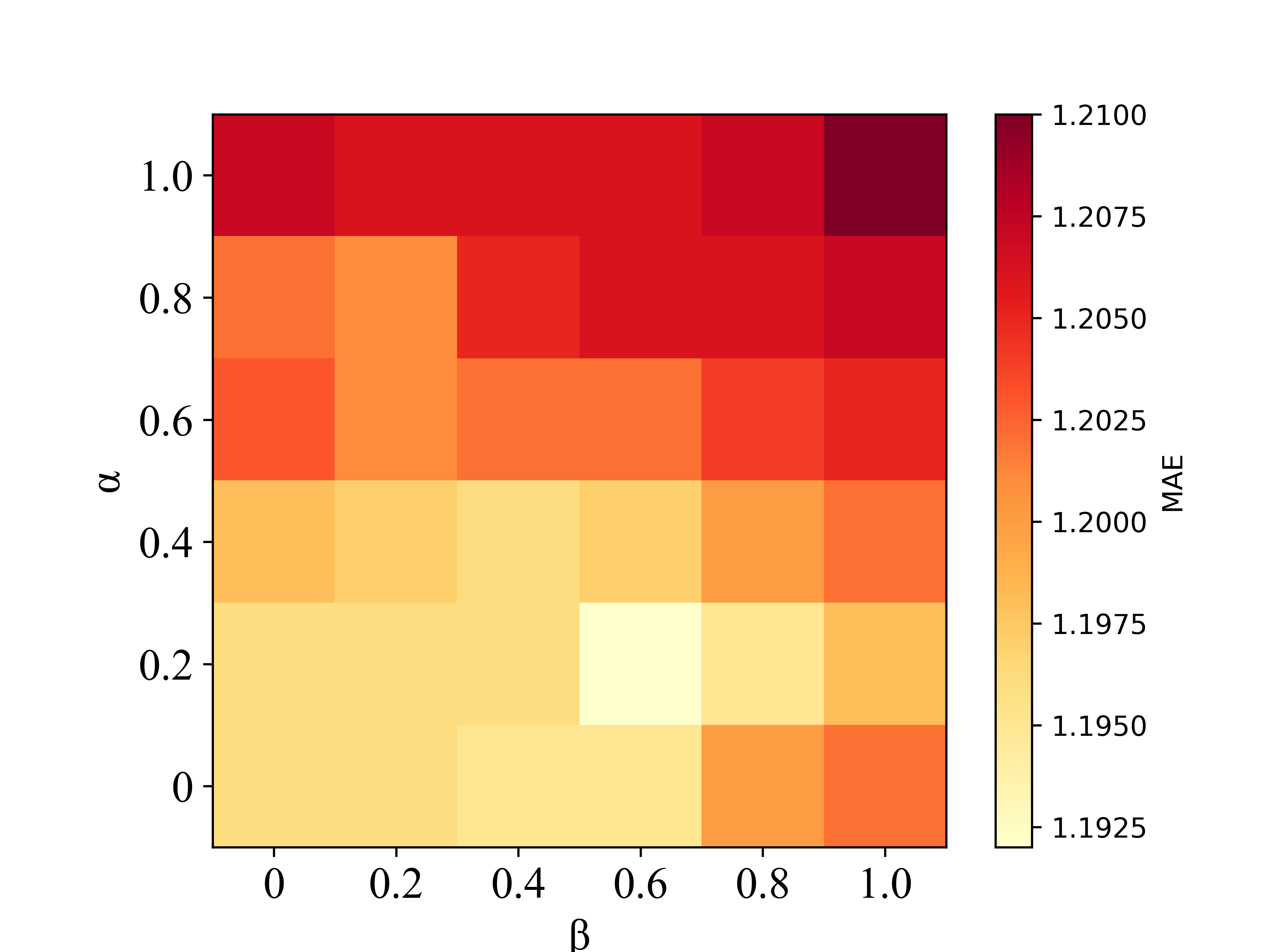}
    \caption{Comparison of MAE (Mean Absolute Error) for different values of residual ratios $\alpha$ (visual) and $\beta$ (textual) in Visual-Language Adapter of DSN. The lighter color of the heat map means lower MAE, i.e. better results. The best result that MAE $=1.192$ is achieved when $\alpha=0.2$ and $\beta=0.6$.}
    \label{fig:residual}
\end{figure} 

\subsubsection{Ablating the encoders of Visual-Language Adapter}
We further verify the effectiveness of using multimodal encoders to extract image and textual features. We choose the state-of-the-art encoders Swintransformer\cite{Swintransformer} and BERT\cite{devlin2018bert} for images and text, respectively, to compare with CLIP used in DSN. To fairly compare the performance of different encoders, we set the residual ratios $\alpha$ and $\beta$ to $1$, i.e., no fine-tuning is adopted. We will verify the effectiveness of our fine-tuning strategy later. Table \ref{tab:VLA} shows the prediction results for different combinations of image and text encoders. We can find that replacing the unimodal encoders with the text or image branch of CLIP, respectively, has some improvement on the results. And encoding both images and text with CLIP achieves significantly better results than using BERT and SwinTransformer. This means that learning the dependency between images and text with a multimodal encoder like CLIP is useful for improving prediction performance.

\begin{table}
    \centering
    \begin{tabular}{lrr}
        \toprule
        Methods & MAE$\downarrow$ & SRC$\uparrow$ \\
        \midrule
        SwinTransformer + BERT & 1.269 & 0.723 \\
        SwinTransformer + CLIP-Text & 1.246 & 0.731 \\
        CLIP-Image + BERT & 1.237 & 0.746 \\
        CLIP-Image + CLIP-Text & \textbf{1.210} & \textbf{0.755} \\
        \bottomrule
    \end{tabular}
    \caption{Comparison with different combinations of image and text encoders. The last line is the encoders used by DSN. The best results are in bold.}
    \label{tab:VLA}
\end{table}

\begin{table}
    \centering
    \begin{tabular}{lrr}
        \toprule
        Methods & MAE$\downarrow$ & SRC$\uparrow$ \\
        \midrule
        First level & 1.281 & 0.724 \\
        Second level & 1.269 & 0.733 \\
        Third level & 1.253 & 0.747 \\
        Concatenation of three levels & 1.249 & 0.752 \\
        Summation of three levels & 1.238 & 0.751 \\ 
        HCE & \textbf{1.192} & \textbf{0.763} \\
        \bottomrule
    \end{tabular}
    \caption{Comparison with different methods to encode three-level category information. HCE denotes the Hierarchical Category Embedding used by DSN. The best results are in bold.}
    \label{tab:HCE}
\end{table}

\begin{table}
    \centering
    \begin{tabular}{lrr}
        \toprule
        Methods & MAE$\downarrow$ & SRC$\uparrow$ \\
        \midrule
        DSN (w/o short-term dependency) & 1.223 & 0.754 \\   
        DSN (w/o long-term dependency) & 1.237 & 0.749 \\
        DSN & \textbf{1.192} & \textbf{0.763} \\
        \bottomrule
    \end{tabular}
    \caption{Ablations on Temporal Fusion of DSN. The long-term dependency denotes the attention mechanism and the short-term dependency is LSTM with the gating layers. The best results are in bold.}
    \label{tab:TF}
\end{table}

\begin{figure*}[!htbp]
    \centering
    \includegraphics[width=0.9\textwidth]{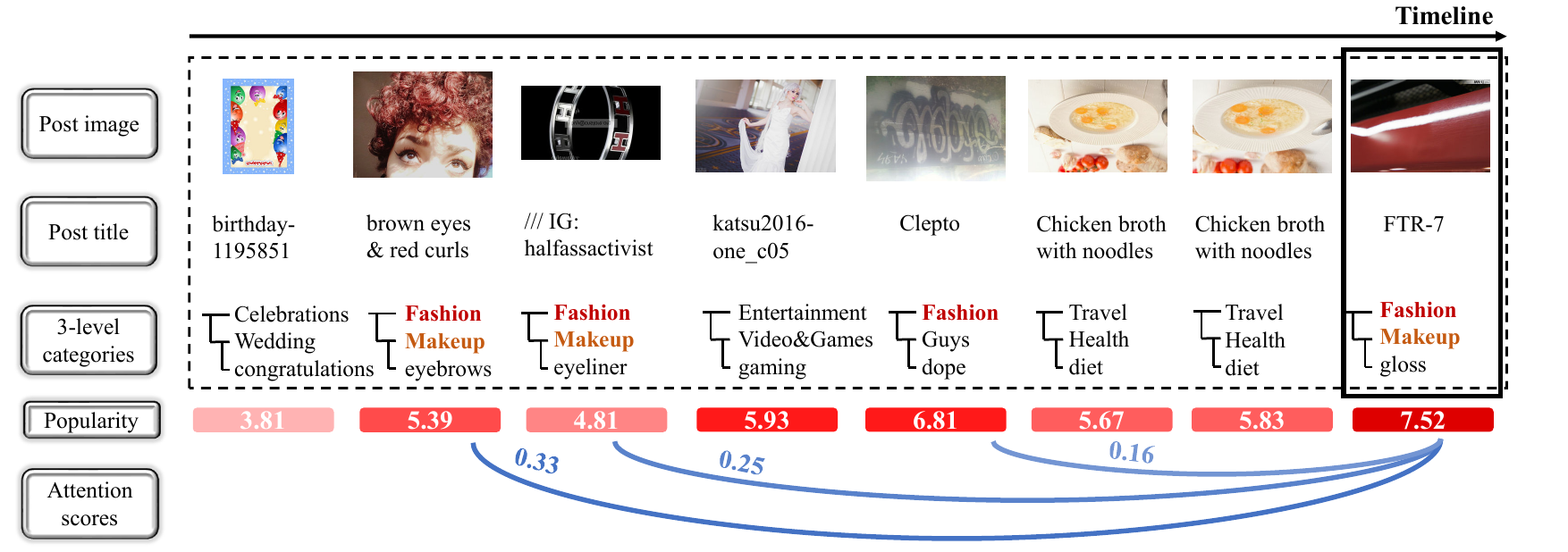}
    \caption{Qualitative results of our DSN when the sequence length $l = 8$. The target predicted post is marked on the far right by the solid box. The ground-truth popularity value is $7.52$ and the prediction result is $7.36$. We show the image, title and category information of each post in the sequence. The ground-truth popularity values and attention scores are also presented, which reflects that posts with similar content share similar popularity. Therefore DSN is able to improve the prediction performance of the target post by aggregating the information of related posts.}
    \label{fig:case}
\end{figure*} 

\subsubsection{Ablating the residual ratios of Visual-Language Adapter} To understand the effectiveness of our fine-tuning strategy in Visual-Language Adapter, we perform an ablation study on the residual ratios $\alpha$ and $\beta$ of visual and textual features. Both ratios take values in the range $\left[0, 1\right]$, and in the experiment, discrete sampling is performed in steps of 0.2. Note that the residual ratios control how much knowledge would be reserved from the pre-trained CLIP model. So when the ratio equals 1, there is no new knowledge is learned, and when the ratio is set to 0, the feature would be fully adapted. Figure \ref{fig:residual} shows the prediction results for different combinations of two ratios. Note that a darker color means a higher MAE, i.e., a worse result. Generally, We can see that when the ratio increases, the MAE also increases at the same time, which means that adjusting the representation as much as possible can produce better results. However, there is an obvious drop in MAE when $\alpha$ increases from 0 to 0.2, which means retaining some of the original knowledge still helps to improve the prediction results. The best result that MAE $=1.192$ is achieved when $\alpha=0.2$ and $\beta=0.6$, respectively, which means compared with text, image features need more fine-tuning to adapt to our task.

\subsubsection{Effectiveness of Hierarchical Category Embedding} We also perform an ablation study on Hierarchical Category Embedding to demonstrate the effectiveness of incorporating the category information. We experiment with different methods to obtain category features, including using category embedding of a single level, directly concatenating three levels together, and the HCE method we proposed in DSN, respectively. From Table \ref{tab:HCE}, compared with only using the category information of one level or simple fusion method (i.e., summation and concatenation), DSN further considers the hierarchical nature of different levels, compressing coarse-grained category information as much as possible while keeping fine-grained category information, which could better model the dependency between different posts and achieve more promising results.

\subsubsection{Effectiveness of Temporal Fusion} We also verify the effectiveness of Temporal Fusion, which aims to model both local and long-term dependencies. The temporal fusion has two main components, LSTM with gating layers for local processing and multi-head attention for long-term dependency. We remove each of these two components separately to evaluate the effectiveness of different dependencies.
Table \ref{tab:TF} shows that removing different components caused different degrees of degradation in prediction performance. When we remove the local processing component, the attention mechanism treats posts as an unordered sequence and the short-term temporal dependency between posts would be lost, so the results get worse. Then we keep the local processing but remove the attention, the model cannot handle long-term dependencies, also leading to worse results.

\subsection{Qualitative Results}
We present a prediction example in the test dataset when the sequence length $l = 8$ in Figure \ref{fig:case}. The target post is on the right highlighted by a solid box. We show the image, title and category information of each post in the sequence. The ground-truth popularity value of the target post is $7.36$ and the prediction result is $7.52$. Note that the popularity values of other posts are used for visualization purposes only and are not used to predict the target posts. We can see the title is a simple description of the content of the image, so mining the correlation of images and text can help DSN better model the content of the post. The categories of the target post are fashion-makeup-gloss. The popularity values and attention scores show that posts with similar content (i.e., images, text and categories) share similar popularity, so DSN could improve prediction performance by integrating posts that have similar content.

\section{Conclusion}
In this work, we present Dependency-aware Sequence Network (DSN), a novel prediction framework for social media popularity prediction. Based on the ability of post content to provide valuable clues for popularity prediction, we jointly model multiple post dependencies for better post representations, leading to significantly better performance compared to competitive baselines on Social Media Popularity Dataset. Extensive ablation studies also show that our proposed (i) visual-language adapter (ii) hierarchical category embedding (iii) adaptive temporal fusion method provide significant contributions to our model’s performance. In the future, we plan to consider the dependency between users (\textit{e.g.} social network graph or information cascade graph) to further improve social media popularity prediction.

\bibliographystyle{ACM-Reference-Format}
\bibliography{main}

\end{document}